\documentclass[a4paper,fleqn,preprint]{cas-sc}
\usepackage{siunitx} 
\usepackage[numbers,sort&compress,square]{natbib}
\usepackage{lineno}
\linenumbers
\usepackage[normalem]{ulem}
\nolinenumbers
\def\tsc#1{\csdef{#1}{\textsc{\lowercase{#1}}\xspace}}
\tsc{WGM}
\tsc{QE}
\tsc{EP}
\tsc{PMS}
\tsc{BEC}
\tsc{DE}

\begin{document}
\let\WriteBookmarks\relax
\def\floatpagepagefraction{1}
\def\textpagefraction{.001}

\shorttitle{Radiation Effects on Surface and Bulk Properties of ATLAS18 Miniature Strip Sensors}

\shortauthors{M. Mikeštíková et~al.}

\title [mode = title]{Radiation Effects on Surface and Bulk Properties of ATLAS18 Silicon Sensors under Low- and High-Dose $\gamma$~Irradiation and Annealing }

\author[ASCR]{M. Mikeštíková}
\cormark[1]
\ead{marcela.mikestikova@cern.ch}

\author[SCIP]{V. Fadeyev}
\author[ASCR]{P. Federičová}
\author[UJP]{P. Gallus}
\author[ASCR]{J. Kozáková}
\author[ASCR]{J. Kroll}
\author[ASCR]{M. Kůtová}
\author[ASCR]{J. Kvasnička}
\author[JSI]{I. Mandić}
\author[ASCR]{P. Tůma}
\author[CNM]{M. Ullan}
\author[KEK]{Y. Unno}
\author[FRB]{I. Zatočilová}

\cortext[cor1]{Corresponding author}


\affiliation[ASCR]{organization={Institute of Physics, Academy of Sciences of the Czech Republic},
            addressline={Na Slovance 1999/2}, 
            city={Prague 8},
            postcode={18200},
            country={Czech Republic}}

\affiliation[SCIP]{organization={Santa Cruz Institute for Particle Physics (SCIPP)},
            addressline={University of California}, 
            city={Santa Cruz},
            postcode={CA 95064},
            country={USA}}
            
\affiliation[UJP]{organization={UJP Praha a.s.},
            addressline={Nad Kamínkou 1345}, 
            city={Prague 5 - Zbraslav},
            postcode={15610},
            country={Czech Republic}}
            
\affiliation[JSI]{organization={Experimental Particle Physics Department, Jožef Stefan Institute},
            addressline={Jamova cesta 39}, 
            city={Ljubljana},
            postcode={SI-1000},
            country={Slovenia}}
            
\affiliation[CNM]{organization={Instituto de Microelectronica de Barcelona (IMB-CNM), CSIC},
            addressline={Campus UAB-Bellaterra},
            city={Barcelona},
            postcode={08193},
            country={Spain}}

\affiliation[KEK]{organization={Institute of Particle and Nuclear Study, High Energy Accelerator Research Organization (KEK)},
            addressline={ 1-1 Oho}, 
            city={Tsukuba, Ibaraki},
            postcode={ 305-0801},
            country={Japan}}
            
\affiliation[FRB]{organization={Physikalisches Institut, Albert-Ludwigs-Universität Freiburg},
            addressline={Hermann-Herder-Straße 3}, 
            city={Freiburg},
            postcode={79104},
            country={Germany}}

\begin{abstract}
Silicon strip detectors for the ATLAS Inner Tracker (ITk) at the HL-LHC must withstand harsh radiation conditions, including fluences of up to $1.6 \times 10^{15}$ 1 MeV $n_\mathrm{eq}/\mathrm{cm}^2$ and total ionizing doses (TID) of up to 66 Mrad.
These requirements are met using radiation-hard $n^+$-in-$p$ technology implemented in the ATLAS18 silicon strip sensors currently under production.
 
This work presents a combined study of $\gamma$-irradiation effects in ATLAS18 silicon sensors, including both segmented miniature strip sensors ("minis") and unsegmented MD8 diodes fabricated on ATLAS18 production wafers. The samples were irradiated with a~$^{60}$Co $\gamma$~source to multiple low TIDs between 0.5 and \SI{100}{krad}, corresponding to the dose range relevant for the early operational phase of the ITk tracker. Additional measurements extending up to a few Mrad were performed to investigate the saturation of surface related damage effects.
 
Post-irradiation characterization included measurements of total, bulk, and surface leakage currents, as well as capacitance-voltage measurements used to extract the full depletion voltage. The thermal stability of radiation-induced defects was studied using isochronal annealing between 80~°C and 300~°C and isothermal annealing at 60~°C and 160~°C. 

In addition, complementary studies of MD8 diodes irradiated to ultra-high doses of several hundred Mrad, well beyond the ATLAS ITk requirements, are included to investigate possible bulk-related effects induced by pure $\gamma$ irradiation and their annealing behaviour.
The combined analysis of low- and ultra-high-dose irradiation provides a comprehensive picture of surface- and bulk-related $\gamma$-induced effects in ATLAS18 silicon sensors and their thermal evolution.

\end{abstract}




\maketitle

\section{Introduction}

Silicon tracking detectors operating at the High-Lumino-sity LHC (HL-LHC) will be exposed to extremely high radiation levels, leading to the formation of electrically active defects that degrade detector performance. Radiation induced damage in silicon devices originates from two primary mechanisms: non-ionizing energy loss (NIEL) and total ionizing dose (TID). NIEL, commonly referred to as bulk damage, arises from the displacement of silicon atoms from their lattice sites, producing point and cluster defects that introduce deep energy levels in the band gap and modify the effective space charge concentration, leakage current, and charge collection properties.

In contrast, ionizing radiation primarily affects the insulating SiO$_2$ layer and the Si-SiO$_2$ interface, resulting in so-called surface damage. Under $\gamma$~irradiation from a~$^{60}$Co~sour-ce, displacement damage is caused mainly by Compton electrons with energies up to $\approx$ 1\,MeV, which generate point defects only. 
Surface damage leads to the formation of interface trap states and the accumulation of a positive oxide charge. These effects can not only induce an inversion layer, degrading the electrical isolation between the strips, but also enhance surface current generation via interface traps acting as Shockley–Read–Hall generation centers \cite{Yoshida:NIMA2003}.

To meet the radiation-level requirements of the \mbox{HL-LHC}, the silicon strip sensors developed for the Inner Tracker (ITk) \cite{TDR} of the ATLAS experiment \cite{ATLAS} are based on radiation-hard $n^+$-in-$p$ technology and fabricated on \mbox{6-inch} wafers with a~physical thickness of $320\,\unit{\um}$. The sensors are currently in the final phase of large-scale production.
Their radiation tolerance was verified through extensive irradiation studies conducted during their development, including exposures to various particle types and energies. 
These studies covered fluences up to the maximum expected value 
for the ITk strip region, located at radial distances of approximately 40 - 100~cm from the beamline and extending over the longitudinal range from z = -300~cm to +300~cm, corresponding to 1.6 $\times$ 10$^{15}\;1\,\mathrm{MeV}\;\mathrm{n_{eq}/cm^2}$ and TID of 66~Mrad anticipated for the ultimate integrated luminosity of 4000 fb$^{-1}$ of HL-LHC operation \cite{Public}.

Previous studies of $\gamma$~irradiated diodes at high TID - up to several hundred Mrad - have shown that the bulk leakage current increases approximately linearly with TID, while surface currents tend to saturate \cite{zatocilova,2024}. However, in these investigations the lowest studied TID values of 66~Mrad were already well above the region in which surface charge saturation is expected. More recent work on $\gamma$~irradiated diodes \cite{2025} explored the low-dose regime and reported no saturation of radiation induced surface charge up to a~TID of 100~krad, suggesting that the saturation threshold lies between 100~krad and several tens of Mrad. 
While these studies separately describe the effects observed at low or high doses, a consistent comparison of $\gamma$-induced surface and bulk effects across a broad TID range, including their annealing behavior, remains limited.

The present work builds upon these earlier studies and extends the investigation to additional dedicated test structures from ATLAS18 production wafers, namely miniature strip sensors ("minis"). In contrast to non-segmented MD8 diodes, minis feature a~segmented geometry representative of the ITk strip layout, enabling a~direct comparison of radiation induced surface and bulk defects in non-segmented and segmented devices. The samples were irradiated with a~$^{60}$Co $\gamma$~source to a~series of low TIDs ranging from 0.5 to 100~krad, covering the dose range most relevant for the early phase of ITk operation. Additional measurements at higher doses, up to a~few Mrad, were performed to investigate the onset and saturation of surface-related effects.
In addition, results from previous ultra-high-dose irradiation studies of MD8 diodes, extending to several hundred Mrad, are included to investigate possible bulk-related effects induced by pure $\gamma$ irradiation and their annealing behaviour.
A~comprehensive post-irradiation electrical characterization was performed, including current-voltage (IV) and capacitance-voltage (CV) measurements as functions of dose, temperature, and annealing history. Particular emphasis was placed on separating bulk and surface leakage current components and on studying the thermal stability of radiation induced defects through isochronal annealing up to 300\,$^\circ\mathrm{C}$ and isothermal temperature annealing at 60\,$^\circ\mathrm{C}$ and 160\,$^\circ\mathrm{C}$. In addition, temperature-dependent measurements were used to extract activation energies for the bulk and surface current components.

\section{Samples and irradiation}

Measurements were performed on $n^+$-in-$p$ MD8 diodes and minis from ATLAS18 production wafers, both with an active thickness of $295\,\unit{\um}$. The MD8 diodes have an active area of 0.545~cm$^2$ and were studied in two variants: regular MD8 devices and MD8-p diodes with an additional $p$-stop implant between the bias and guard rings to improve isolation between the active diode area and the edge region and reduce edge-related surface currents \cite{UnnoSpecs}. The minis have an active area of 0.64~cm$^2$ with 104 strips \cite{UnnoA17}.

\begin{figure}
    \centering
    \includegraphics[width=0.4\linewidth]{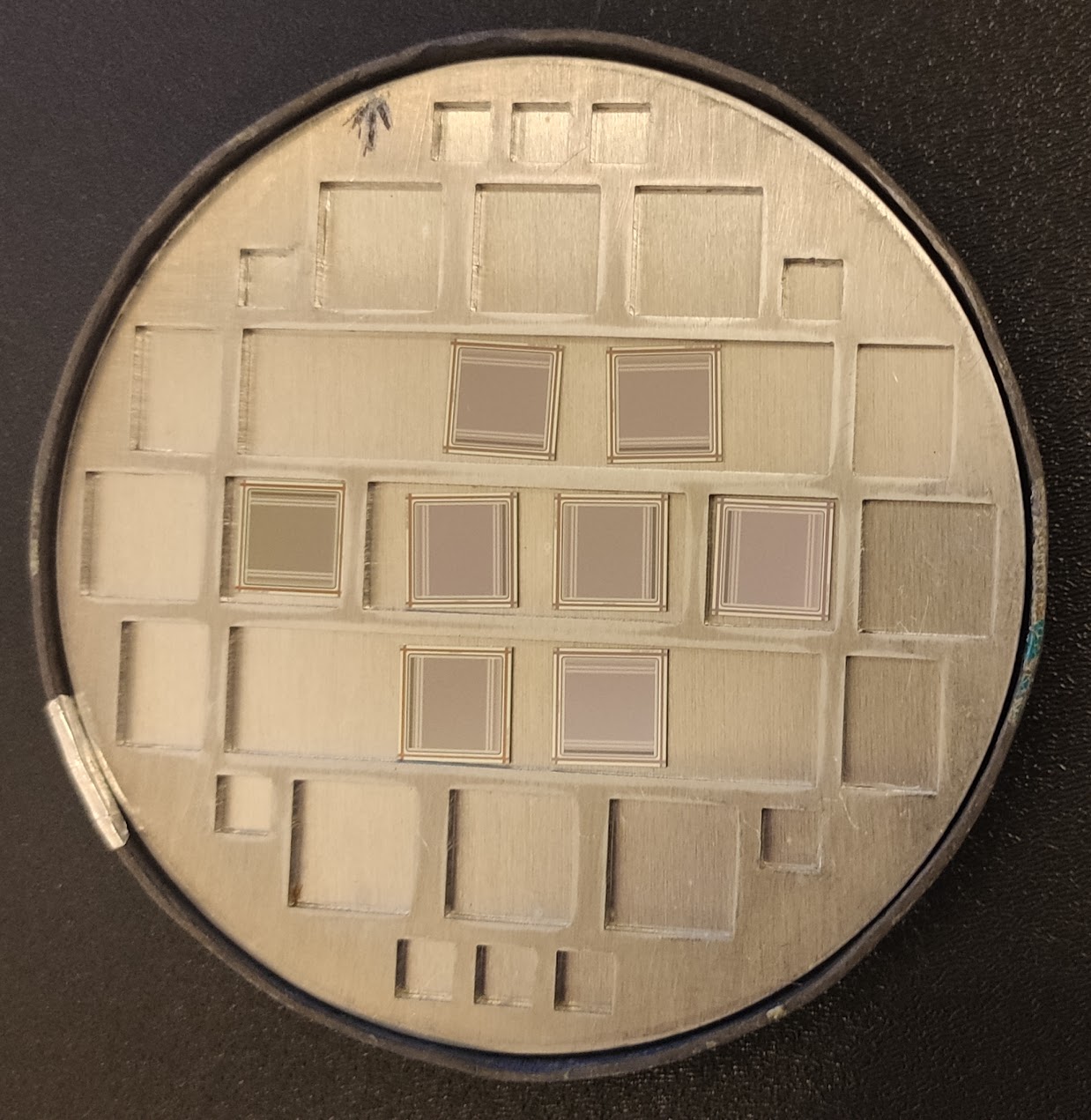}
    \caption{The miniature sensors in CPE box.}
    \label{fig:Fig1}
\end{figure}

All samples were irradiated with $^{60}$Co $\gamma$~rays at UJP Praha a.s. \cite{UJP} to a~total of 14 different TID levels spanning the range from 0.5 to 100~krad. To extend the study toward higher doses and to investigate the saturation behaviour of surface-related effects, additional measurements were performed up to TIDs of 3~Mrad for minis and 24~Mrad for diodes. The irradiations were carried out with dose rates \mbox{1.6 - 8.5~krad/min}, with an estimated dose rate uncertainty below $5\,\%$.
During irradiation, the samples were placed inside a~charge particle equilibrium (CPE) box composed of aluminium and lead, in accordance with ESA/SCC recommendations \cite{ESA} for $\gamma$~irradiation of semiconductor devices. The minis in CPE box are shown in Fig. \ref{fig:Fig1}. This configuration establishes charge particle equilibrium, minimizes dose enhancement effects, and ensures uniform energy deposition in the silicon. The sample temperature was maintained below 35\,$^\circ\mathrm{C}$ using air fan throughout irradiation. Following irradiation, the samples were immediately stored at temperatures below \mbox{-20}\,$^\circ\mathrm{C}$ to suppress uncontrolled annealing prior to electrical measurements.

\section{Experimental methods and results}

To investigate radiation induced effects caused by $^{60}$Co $\gamma$~irradiation, a~comprehensive electrical characterization was performed on MD8 diodes and minis before and after irradiation. The experimental program included IV and CV measurements, annealing studies, temperature-dependent measurements, and transient current technique (TCT)
analy\-ses.

\subsection{IV measurements with separation of bulk and surface currents} \label{sec:current}

Special care was taken to determine individual leakage current components, namely the bulk current ($I_{\mathrm{bulk}}$) flowing exclusively through the active volume of the diode or strip sensor bounded by the guard ring, and the surface-related current ($I_{\mathrm{surf}}$), which dominates outside the guard ring region.
The total leakage current ($I_{\mathrm{tot}}$) is defined as
\begin{equation*}
    I_{\mathrm{tot}} = I_{\mathrm{bulk}} + I_{\mathrm{surf}}.
\end{equation*}

\begin{figure}
                \centering
                \includegraphics[width=0.4\linewidth]{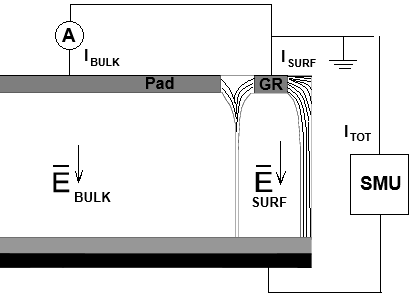}
                \caption{Schematic of the test configuration of $I_{\mathrm{tot}}$ and $I_{\mathrm{bulk}}$ in diodes with a~contactable guard ring.}
                   \label{fig:Fig2a}
\end{figure}

For MD8 diodes equipped with a~contactable guard ring, $I_{\mathrm{bulk}}$ and $I_{\mathrm{surf}}$ were measured directly by independently monitoring the currents collected at the diode pad and at the guard ring, respectively. A~schematic of the measurement configuration is shown in Fig. \ref{fig:Fig2a}. This setup enables a~clear separation between currents generated in the depleted bulk and those originating from the surface and edge regions. More information about the current-generation mechanisms at the sensor periphery from TCAD simulation can be found in Ref. \cite{Peilin}.

\begin{figure}
                \centering
                \includegraphics[width=0.7\linewidth]{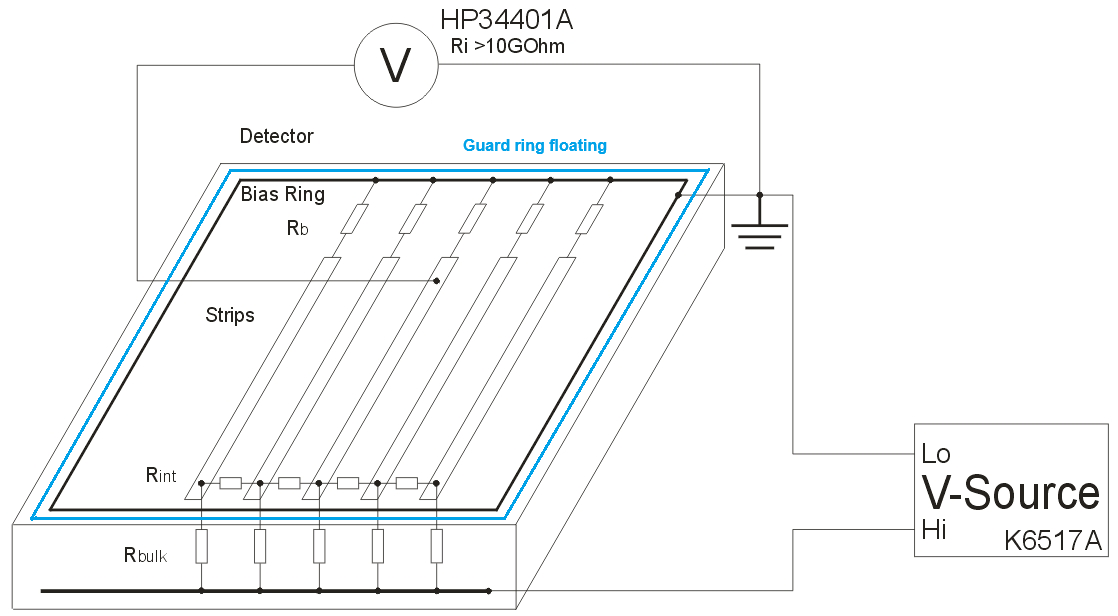}
                \caption{Schematic of the test configuration of $I_{\mathrm{tot}}$ and $I_{\mathrm{bulk}}$ in strip sensors with non-contactable guard ring.}
                   \label{fig:Fig2b}
\end{figure}

For minis, where the guard ring is not contactable, the bulk current was extracted indirectly using the method illustrated in Fig. \ref{fig:Fig2b}. 
The current flowing through an individual strip ($I_{\mathrm{strip}}$) was determined from the voltage drop across the corresponding bias resistor, measured using a~high input resistance voltmeter ($R_{\text{input}}>10\,\mathrm{G\Omega}$). The bias resistance and voltage drop were measured on six randomly selected strips as well as on the first three strips adjacent to the sensor edge, yielding a~highly uniform average bias resistance of
$R_{\text{bias}}=(1.584 \pm 0.002)\,\mathrm{M\Omega}$.
This uniformity confirms that compensating interstrip current through the interstrip resistance is negligible and that the voltage drop can be assumed identical for all strips. Under these conditions, the bulk current was calculated as $I_{\text{bulk}} = n~ \times\, I_{\text{strip}},$
where \mbox{$n = 104$} is the number of strips in the mini sensor. The surface current was then determined as the difference between the total current and the bulk current, $I_{\text{surf}} = I_{\text{tot}}-I_{\text{bulk}}$.
Owing to the large ratio between the voltmeter input resistance and the bias resistance, the influence of the measurement setup on the current distribution in the sensor was negligible.

All IV measurements were carried out on a~probe station under controlled environmental conditions, with the temperature stabilized at $(20 \pm 0.1)\,^\circ\mathrm{C}$ and relative humidity maintained below $1\,\%$ for both unirradiated and irradiated samples.

Prior to irradiation, bulk and surface leakage currents in MD8 diodes are comparable in magnitude, with the surface component being slightly larger in devices equipped with a~$p$-stop implant between the bias and the guard rings, as reported in previous study \cite{2025}.

\begin{figure}
                    \centering
                    \includegraphics[width=0.7\linewidth]{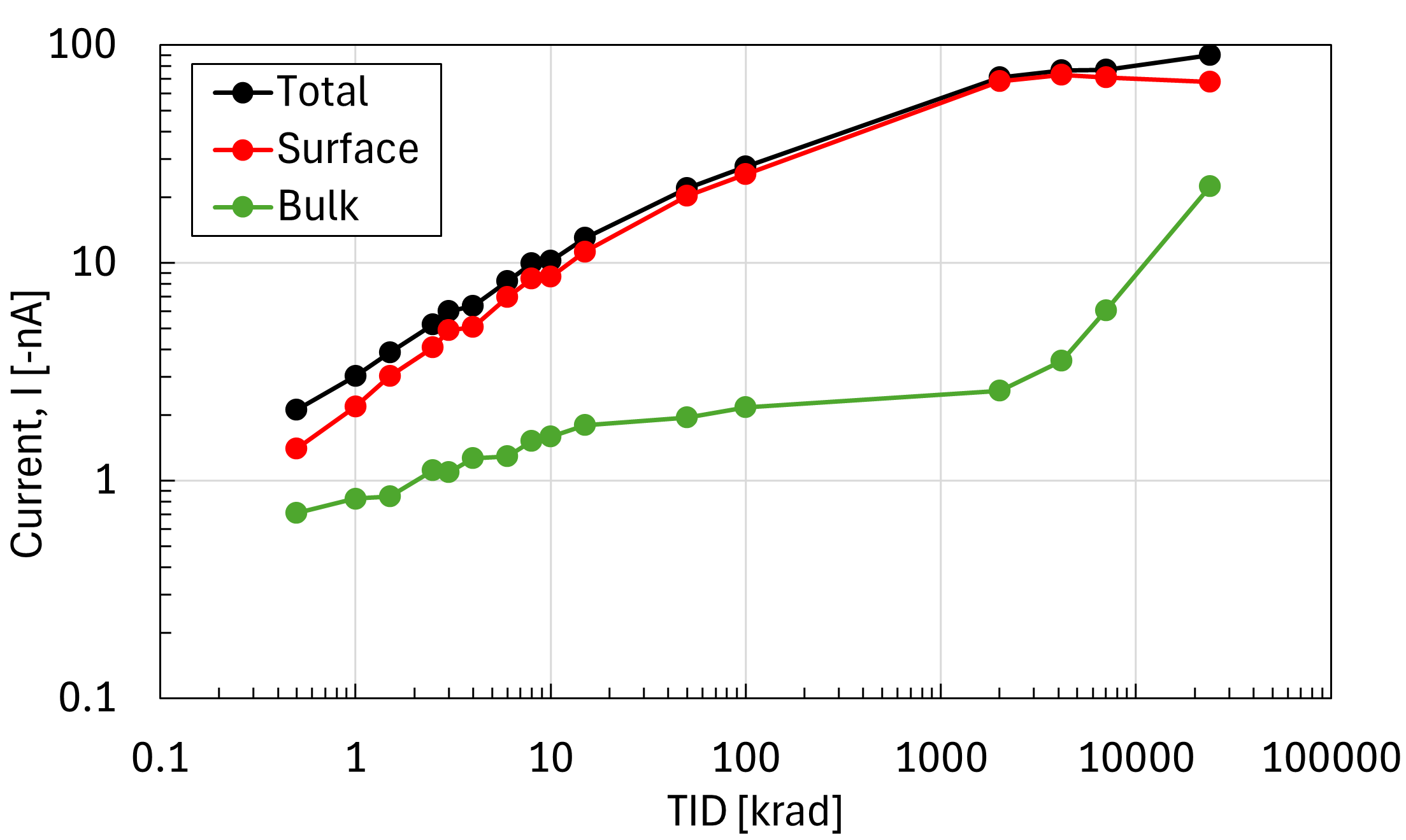}
                                \caption{Evolution of total, bulk and surface currents in \mbox{MD8-p} diodes as a~function of TID, measured at $V_{\mathrm{bias}}$=~-300~V.}
                   \label{fig:Fig4_new}
\end{figure}

Fig. \ref{fig:Fig4_new} shows the evolution of total, bulk, and surface leakage currents in MD8-p diodes as a~function of TID for low $\gamma$~irradiation levels, measured at a~bias voltage \mbox{$V_{\mathrm{bias}}$ = -300~V}. After irradiation, a~clear increase of the total leakage current is observed even at the lowest investigated TID values. This increase is dominated by the surface current (which is a~sum of the real surface current and the bulk generation current in the edge volume), while the bulk current remains significantly smaller over most of the low dose range. The surface current shows signs of saturation at a~TID of approximately 2~Mrad. This behaviour is consistent with the accumulation and eventual saturation of radiation induced charges at the Si-SiO$_2$ interface. In contrast, the bulk current exhibits a~much weaker dependence on TID in this regime, remaining nearly constant up to about 2~Mrad and increasing only at higher doses. 
 This behaviour is consistent with the expected linear dependence of the bulk current on TID observed at higher doses, while the apparent plateau at low TID is likely influenced by the limited sensitivity of the measurement at very low current levels.

These results indicate that, for low $\gamma$~doses relevant to the early phase of ITk operation, the leakage current increase in MD8 diodes is primarily driven by surface-related effects, while bulk damage remains limited.

\begin{figure}
                \centering
                \includegraphics[width=0.7\linewidth]{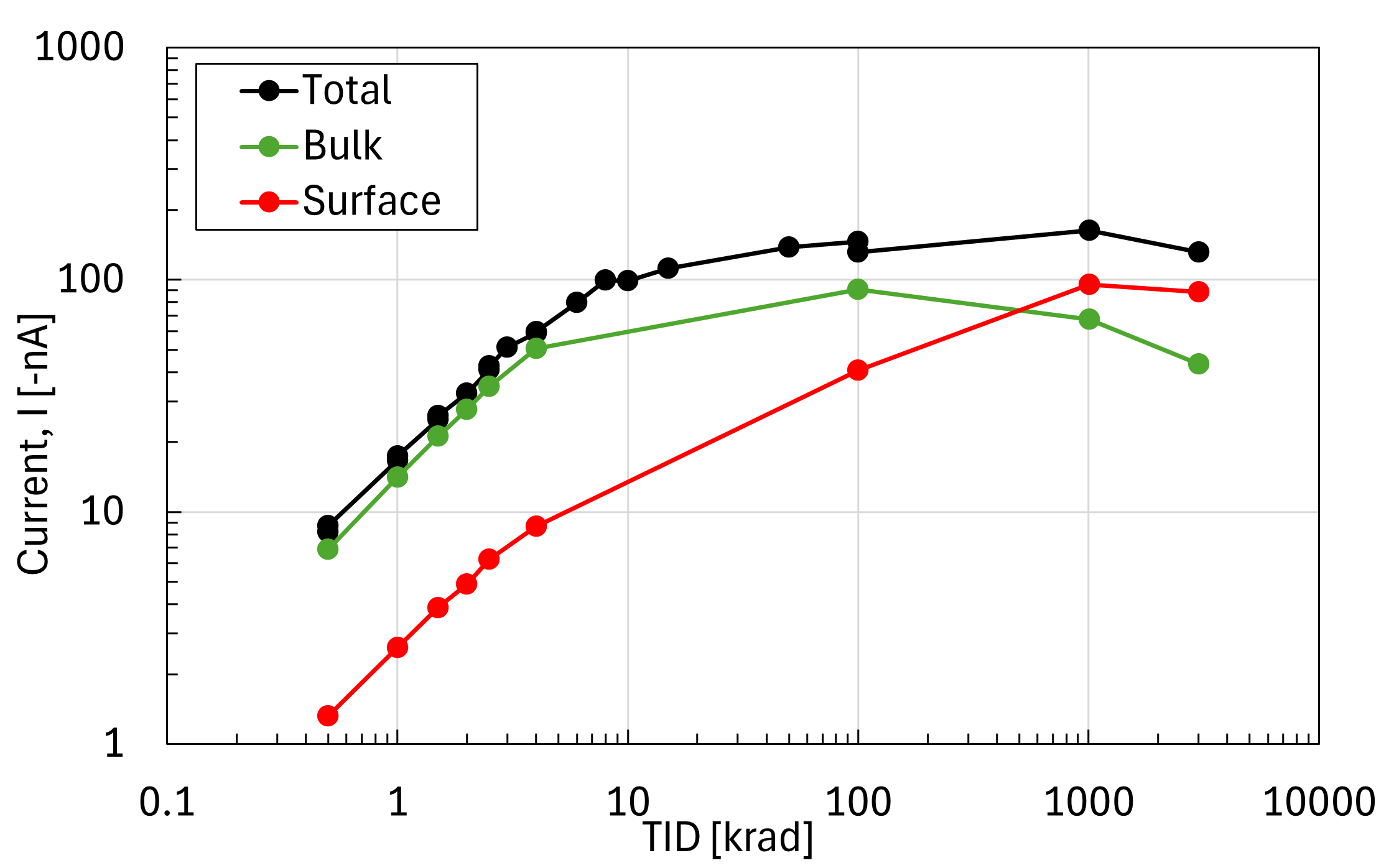}
                \caption{Evolution of total, bulk and surface currents in minis as a~function of TID. $V_{\mathrm{bias}}$ =~-300~V.}
                   \label{fig:Fig5_new}
\end{figure}

Fig. \ref{fig:Fig5_new} shows the total, bulk, and surface leakage currents measured in mini sensors as a~function of TID. 
Unlike in diodes, in minis the bulk current dominates the total current at the very lowest doses; however, as the dose increases, the surface component becomes dominant, similarly to the behavior observed in diodes. The origin of the bulk current decrease at the upper end of the investigated low TID range requires further study.

Figures \ref{fig:Fig4_new} and \ref{fig:Fig5_new} show that the bulk current of the minis is about 40 times higher than that of the MD8 diodes at a TID of 100~krad. This behavior can be explained by the segmented geometry of the minis. The bulk current inside the guard ring consists of a bulk component generated in the space-charge region and a surface component generated at interface traps between the strips. Owing to the much larger total interface area in the mini sensors compared to that in the edge region of the MD8 diode, a significantly higher current is observed in the minis. 
For full size strip sensors, although the current inside the guard ring is expected to be even larger than that outside due to the relatively larger area inside the guard ring compared with mini sensors, the ratio of the bulk component to the surface component in the inside current remains constant, as does its evolution with TID.

As in diodes, indications of surface current saturation are observed around 1 - 2~Mrad.

\subsection{Annealing studies} \label{sec:annealing}

To investigate the thermal stability of radiation induced defects, isothermal and isochronal annealing studies were performed. All irradiated samples underwent a~standard annealing step at 60\,$^\circ\mathrm{C}$ for 80 minutes prior to detailed characterization. In addition, selected samples were subjected to extended annealing at higher temperatures.

\begin{figure}
                \centering
                \includegraphics[width=0.7\linewidth]{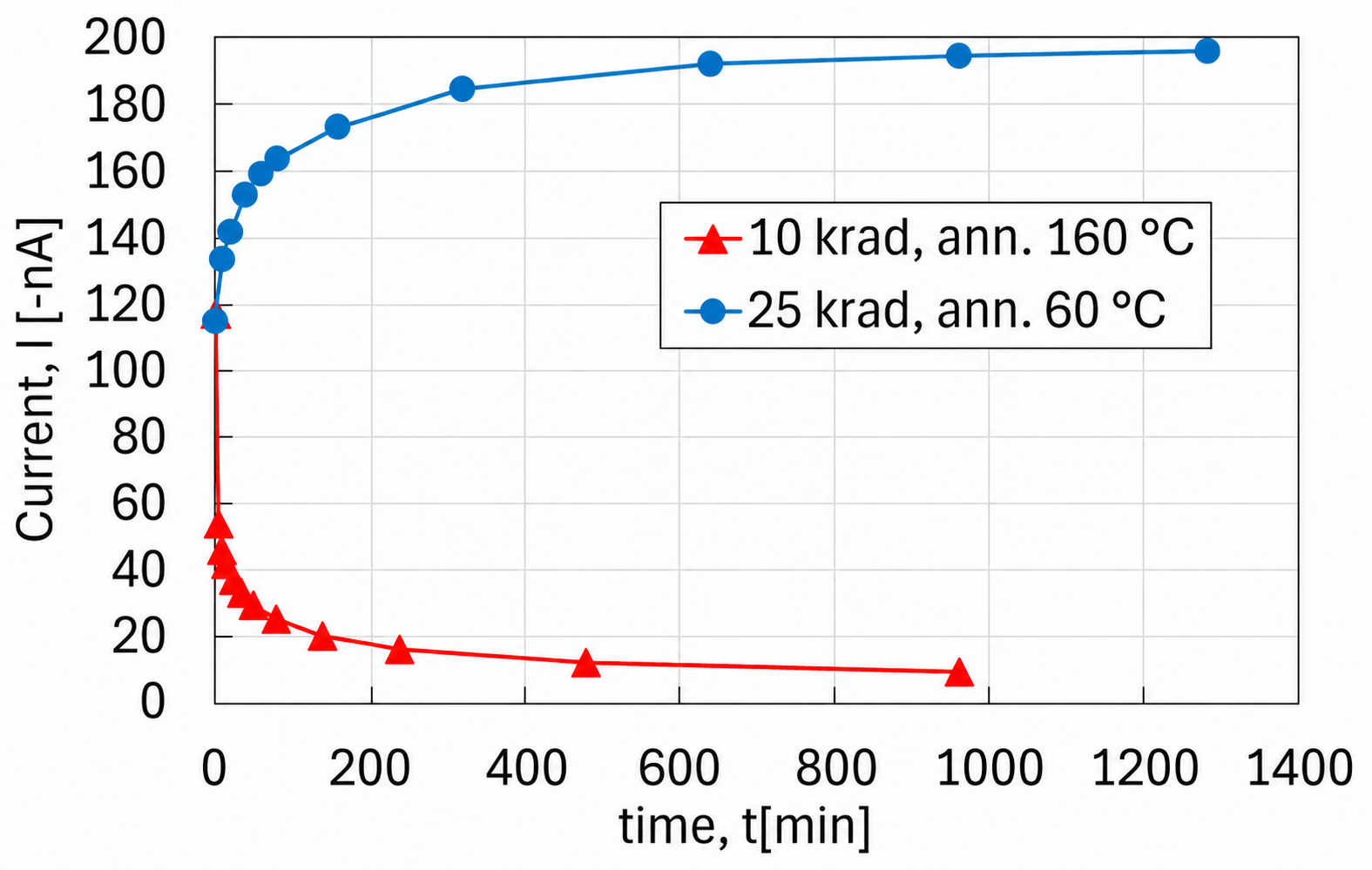}
                \caption{Total current versus annealing time for minis irradiated to 25~krad (60\,$^\circ\mathrm{C}$) and 10~krad (160\,$^\circ\mathrm{C}$); $V_{\text{bias}}$ =~-500~V.}

                                        \label{fig:Fig6}
\end{figure}

Isothermal temperature annealing was performed at 60\,$^\circ\mathrm{C}$ and 160\,$^\circ\mathrm{C}$ for mini sensors irradiated to 25 and 10~krad, respectively. The evolution of the total current as a~function of annealing time is shown in Fig. \ref{fig:Fig6}. At 60\,$^\circ\mathrm{C}$, the total current increases with annealing time, reaching a~maximum after approximately 640~minutes and subsequently saturating, which is characteristic of shallow-trap behavior. In contrast, annealing at 160\,$^\circ\mathrm{C}$ leads to a~strong reduction of the total current, approaching pre-irradiation values. This indicates effective annealing of radiation induced surface defects at elevated temperature and is consistent with previous studies on MD8 diodes \cite{2025}. The CV characteristics remain unchanged after annealing, further confirming that the low TID damage is purely surface-related.

\begin{figure}
                \centering
                \includegraphics[width=0.7\linewidth]{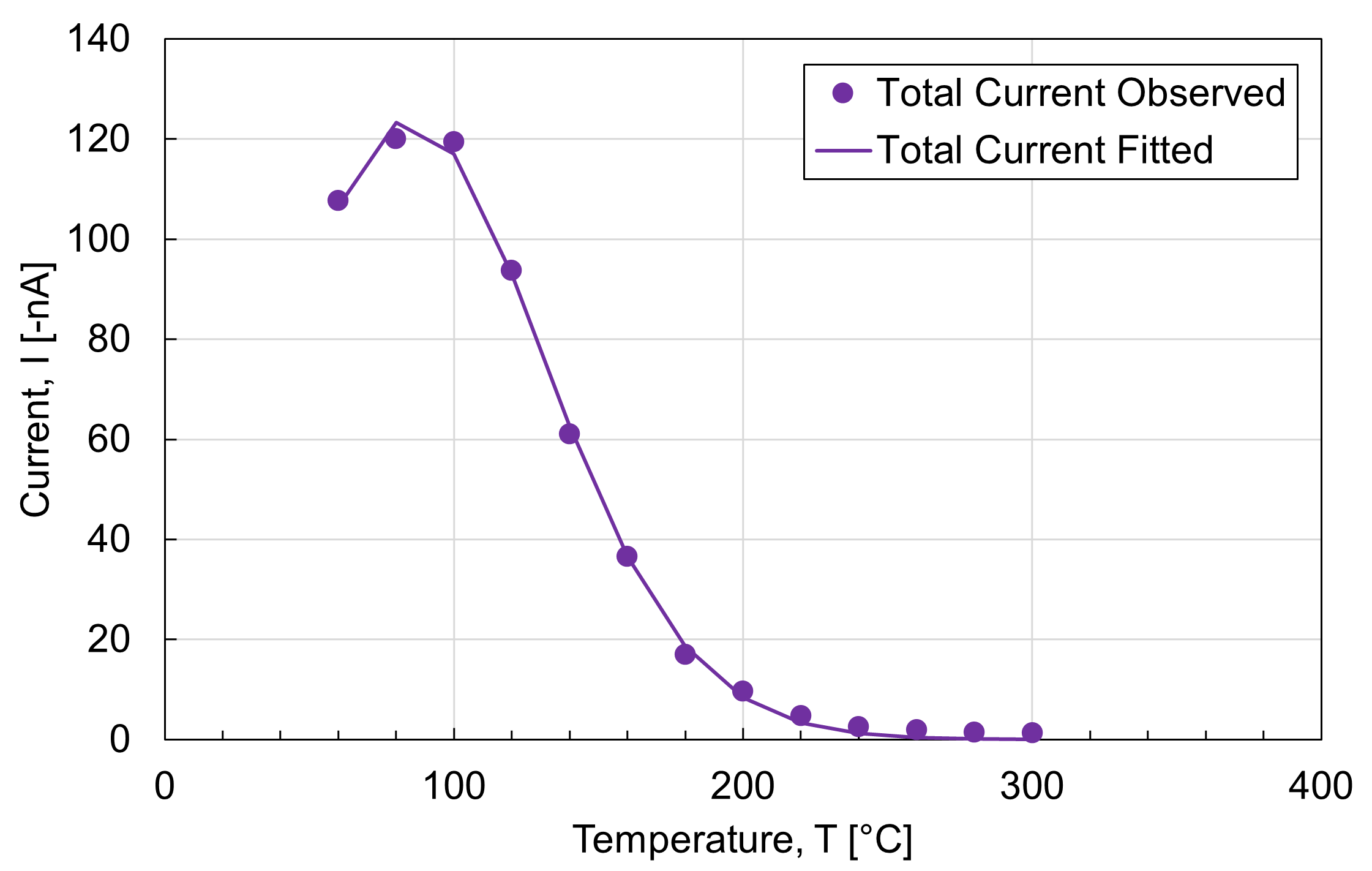}

                                \caption{Evolution of total current as a~function of annealing temperature for mini irradiated to 15~krad. $V_{\text{bias}}$ =~-300~V. Data were  fitted by Eq. ~\eqref{eq:fit_temperature} }
                   \label{fig:Fig7_new}
\end{figure}

Isochronal annealing was carried out in 20-minute steps at temperatures ranging from 60\,$^\circ\mathrm{C}$ up to 300\,$^\circ\mathrm{C}$ for the mini sensor irradiated to a~low $\gamma$~TID of 15~krad. The evolution of the total current as a~function of annealing temperature is presented in Fig. \ref{fig:Fig7_new}. The current shows an increase in the temperature range of approximately \mbox{80 - 100\,$^\circ\mathrm{C}$}. At higher annealing temperatures, the current decreases significantly, returning to values comparable to those before irradiation. This is in agreement with our previous study on MD8 diodes irradiated to the same TID \cite{2025}. For comparison, Ref. \cite{Liao} reports that in $\gamma$~irradiated $n^+$-in-$p$ diodes exposed to TIDs of \mbox{100 - 200~Mrad}, the leakage current remains unchanged up to 200\,$^\circ\mathrm{C}$, rises rapidly between 200\,$^\circ\mathrm{C}$ and 260\,$^\circ\mathrm{C}$, and then decreases strongly at higher annealing temperatures up to 300\,$^\circ\mathrm{C}$.

To extract the activation energy $E$ of the annealing process, the temperature dependence of the current was fitted using an Arrhenius-type relation. The current after annealing at temperature $T_n$ is related to that at the previous temperature step $T_{n-1}$ according to
\begin{equation}
I(T_n) = I(T_{n-1}) \exp\!\left[
-\frac{E}{k_B}
\left(
\frac{1}{T_{\mathrm{ref}}} - \frac{1}{T_n}
\right)
\right],
\label{eq:fit_temperature}
\end{equation}
where $E$ is the activation energy of the annealing process, $k_B$ is the Boltzmann constant, and $T_{ref}$ is a~reference temperature corresponding to the maximum observed current. The same fitting procedure was applied to both minis and MD8 diodes, yielding consistent results and confirming a~common annealing mechanism. 
The extracted activation energies obtained from the fits are summarized in \mbox{Table \ref{tab:activation}}. For the MD8 diode, the activation energy derived from the total current is \mbox{$E_{tot} = 0.064$~eV}, in good agreement with the value obtained for the surface current component \mbox{($E_{surf} = 0.065$~eV)}, while the bulk current exhibits a~lower activation energy of \mbox{$E_{bulk} = 0.041$~eV}. This confirms that the annealing behavior of the total current in the MD8 diode is dominated by surface-related effects associated with radiation induced oxide and Si-SiO$_2$ interface states at low TID. For the segmented mini, a~higher activation energy of $E = 0.113$~eV is extracted from the total current, indicating a~stronger temperature dependence of the annealing process. Despite the quantitative differences, the results for both device types are consistent with a~common surface driven annealing mechanism at low TID.

\begin{table}[h]
\caption{Activation energy values of the annealing process for MD8 diode and mini irradiated to 15~krad.}
\centering
\begin{tabular}{l|ccc|c}
\hline
 & \multicolumn{3}{c|}{MD8 diode} & Mini \\
 & Total C& Bulk C& Surface C& Total Current\\
\hline
$E\,[\mathrm{eV}]$ & 0.064 & 0.041 & 0.065 & 0.113 \\
$T_{ref}\,[^\circ\mathrm{C}]$ & 100.85 & 104.58 & 99.51 & 94.56 \\
\hline
\end{tabular}
\label{tab:activation}
\end{table}
The effect of high temperature annealing was further investigated using MD8 diodes irradiated to TID of 15~krad and 630~Mrad in order to compare the annealing behavior after low and high $\gamma$~irradiation. The diodes were annealed at 300\,$^\circ\mathrm{C}$ for 1~hour, and the resulting currents were analyzed. 
 
\begin{figure}

                                \centering
                \includegraphics[width=0.7\linewidth]{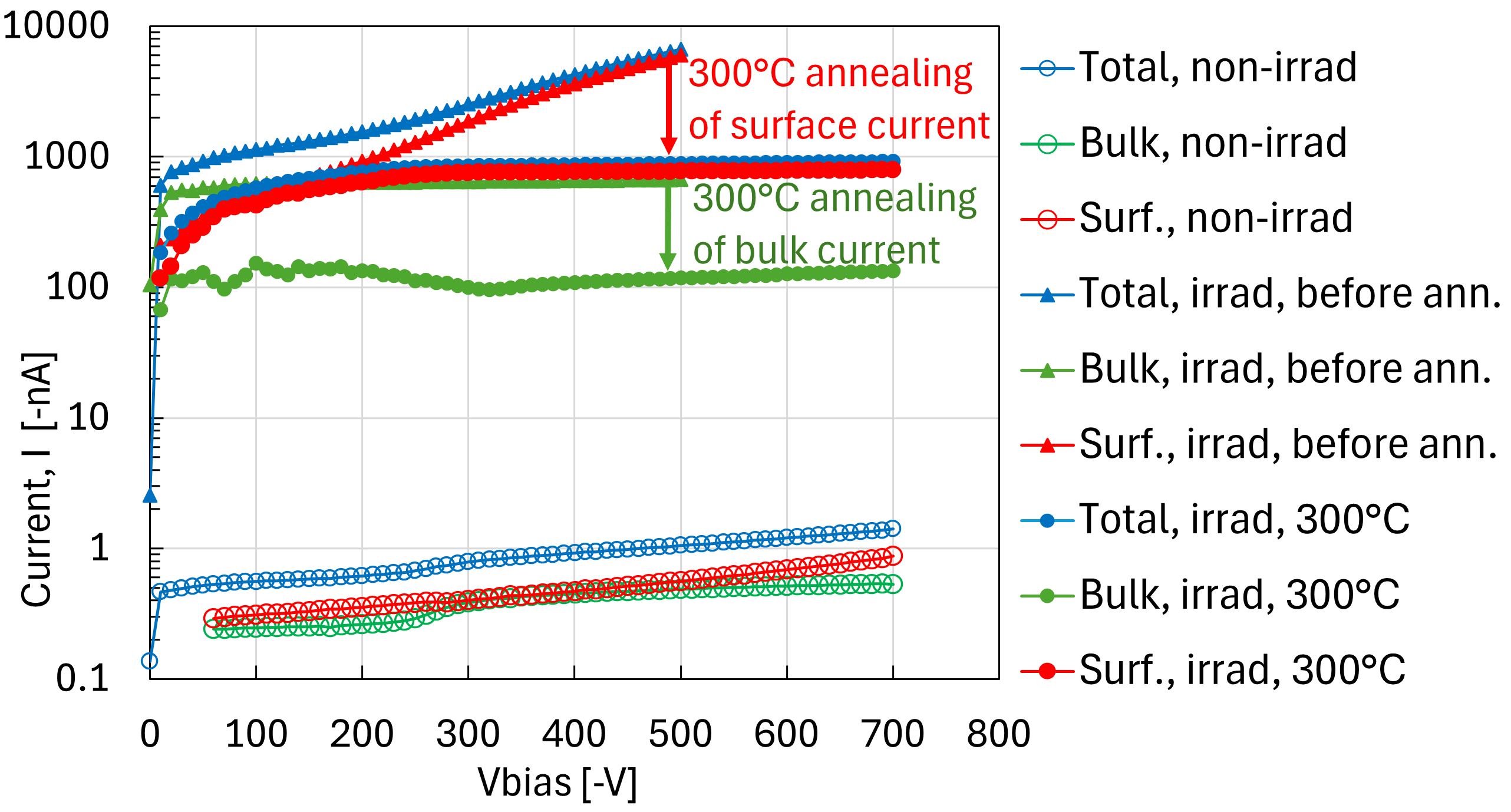}
                \caption{Total, surface, and bulk currents for MD8 diode irradiated to a~high TID of 630~Mrad measured before and after annealing for 1~hour at 300\,$^\circ\mathrm{C}$ (“before annealing” denotes before the 300 °C high-temperature annealing; standard annealing at 60\,$^\circ\mathrm{C}$ for 80~minutes was applied in all irradiated cases). Non-irradiated currents are shown for comparison.}
                   \label{fig:Fig8_new}
\end{figure}

Fig. \ref{fig:Fig8_new} shows the total, bulk, and surface currents of MD8 diodes irradiated to a~high TID of 630~Mrad; corresponding results for low TID irradiation are discussed in Ref. [6]. As in the low TID case, the surface current is the dominant contribution to the total current.
For the low TID diodes, both bulk and surface current components recover to values comparable to those before irradiation after annealing at 300\,$^\circ\mathrm{C}$, indicating a~near-complete recovery of radiation induced effects. In contrast, for high TID diodes, annealing substantially reduces all current components, but they remain above their pre-irradiation values under the applied annealing conditions. This residual current may reflect contributions from radiation induced bulk and surface defects with higher thermal stability. The physical mechanism responsible for this residual current and its temperature dependence is not yet fully understood and is the subject of ongoing investigation.

\begin{figure}
                \centering
                \includegraphics[width=0.7\linewidth]{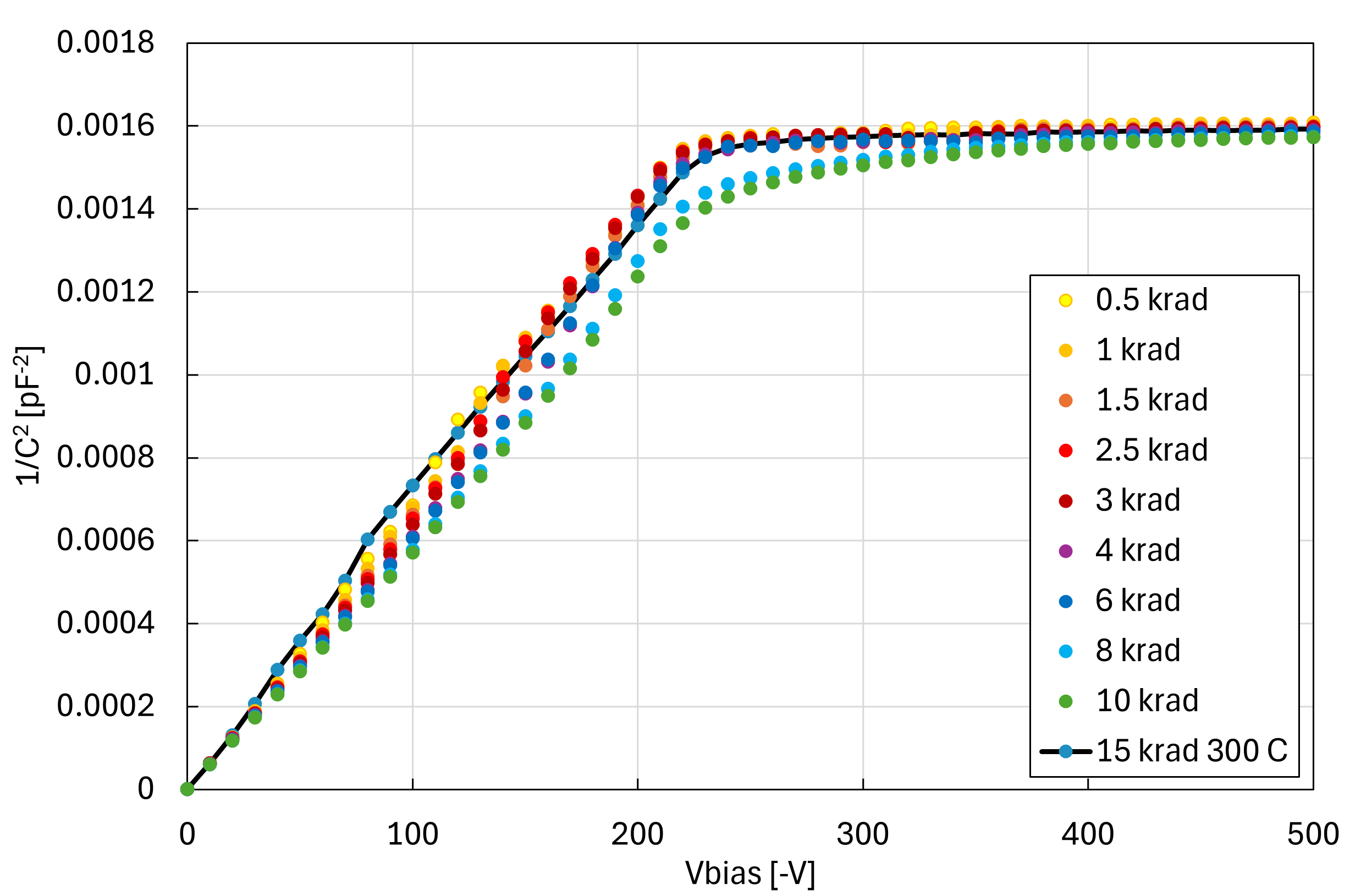}
                \caption{CV characteristics of minis irradiated to low TIDs up to 15~krad, measured before annealing and after 1~hour of annealing at  300\,$^\circ\mathrm{C}$.}
                    \label{fig:Fig9_newa}
\end{figure}

To assess the impact of bulk damage effects, CV characteristics were measured for MD8 diodes irradiated to a~high TID of 630~Mrad and for mini sensors irradiated to low TIDs up to 15~krad. Measurements were performed after standard annealing for 80 minutes at 60\,$^\circ\mathrm{C}$ and after additional annealing for 1 hour at temperatures up to 300\,$^\circ\mathrm{C}$. Low-TID irradiation of mini sensors and high-temperature annealing leave the full depletion voltage ($V_\mathrm{FD}$) unchanged (see Fig. \ref{fig:Fig9_newa}), indicating that low TID irradiation does not induce measurable bulk damage and that the radiation effects are confined to the surface.

\begin{figure}
                \centering
                \includegraphics[width=0.7\linewidth]{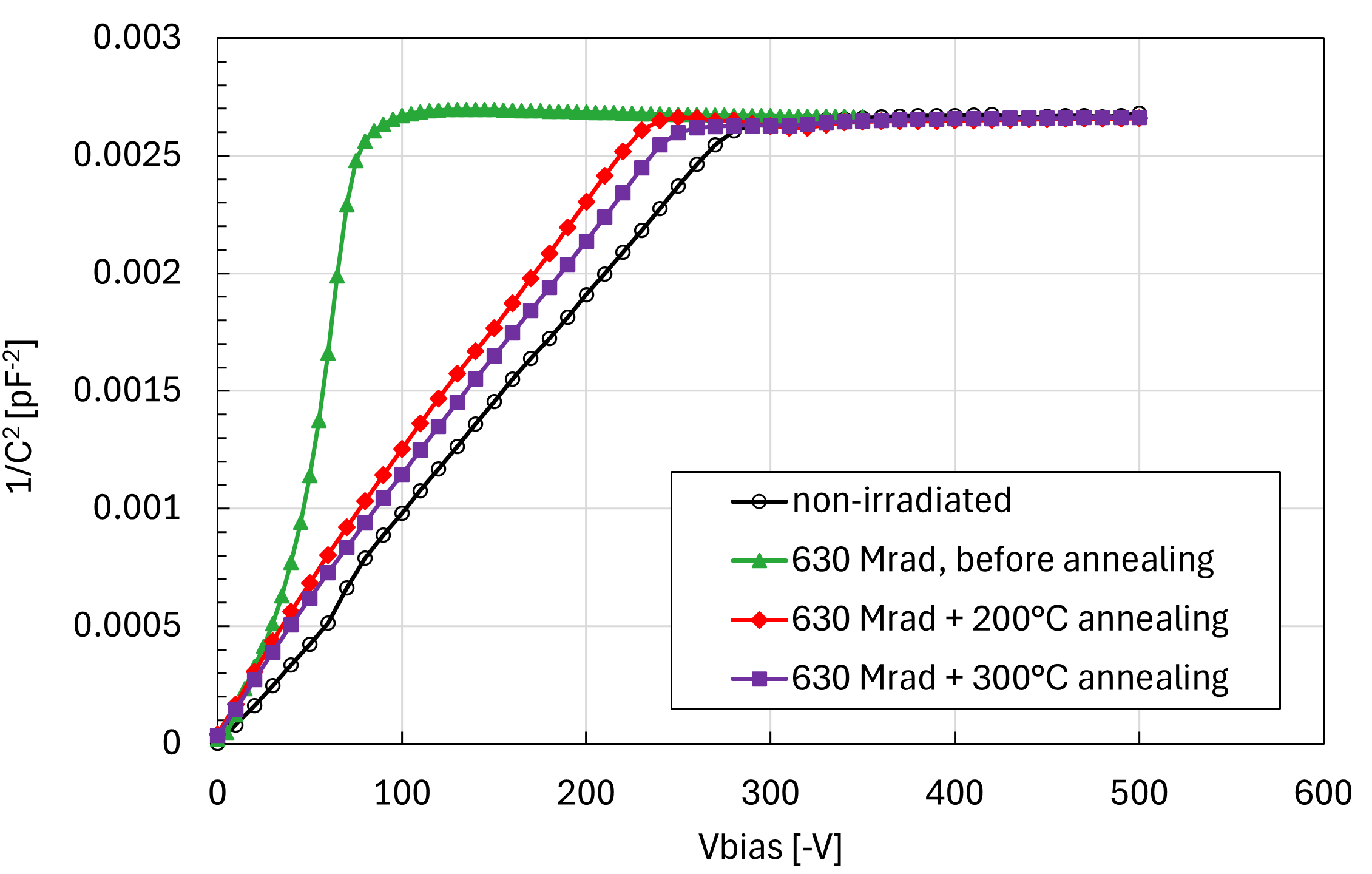}
                \caption{CV characteristics of an MD8 diode irradiated to a~high TID of 630~Mrad, measured before annealing and after 1~hour of annealing at 200\,$^\circ\mathrm{C}$ and 300\,$^\circ\mathrm{C}$ (“before annealing” denotes before the  high-temperature annealing; standard annealing at 60\,$^\circ\mathrm{C}$ for 80~minutes was applied in all irradiated cases). For reference, the CV curve of a~non-irradiated diode is also shown, corresponding to $V_{\mathrm{FD}}$ =~-275 V.}
                    \label{fig:Fig9_newc}
\end{figure}

Fig. \ref{fig:Fig9_newc} shows the CV characteristics of an MD8 diode irradiated to a~high TID of 630~Mrad, measured before annealing and after 1~hour of annealing at 200\,$^\circ\mathrm{C}$ and 300\,$^\circ\mathrm{C}$. For reference, the CV curve of a~non-irradiated diode is also shown, corresponding to $V_{\mathrm{FD}}$ =~-275 V. High TID irradiation leads to a~pronounced reduction of $V_{\mathrm{FD}}$ to $\approx~-70\,\mathrm{V}$, indicating significant radiation induced bulk damage, in agreement with our previous study \cite{2025}. Subsequent high temperature annealing results in a~substantial recovery of the bulk properties, with $V_{\mathrm{FD}}$ increasing to $\approx -250\,\mathrm{V}$ after annealing at 300\,$^\circ\mathrm{C}$, close to the pre-irradiation value.

\subsection{TCT studies} \label{sec:TCT}

The  transient current technique (TCT) was employed to investigate the electric field distribution, the effective space charge concentration, and the $V_{\mathrm{FD}}$ in MD8 diodes after high $\gamma$~irradiation and subsequent high temperature annealing. The measurements were made with the system produced by Particulars \cite{TCT}. \mbox{Top-TCT} measurements were performed using a~red laser with a~wavelength of 660~nm, illuminating the devices from the $n^{+}$ side.

TCT measurements were performed on MD8 diodes irradiated to a~TID of 630~Mrad and annealed at 300~$^\circ$C, and the results are compared with previously reported data~for diodes irradiated to lower (50~Mrad) and higher (828~Mrad) $\gamma$~doses from Ref.~\cite{2024}. Diodes irradiated to 50~Mrad exhibit a~standard $n^+$-in-$p$ electric field configuration with a~full depletion voltage from -250 to -300~V.
In contrast, at higher irradiation levels of 828~Mrad, a~double-junction electric field structure is observed, with a~$p$-type region near the $n^{+}$ side and an $n$-type region near the $p^{+}$ side, separated by a~quasi-neutral bulk. This behavior reflects a~space-charge redistribution caused by carrier trapping at radiation-induced deep levels at high gamma doses and
is accompanied by a~strong reduction of the depletion voltage to approximately -30 to -40 V. The observed decrease of $V_{\mathrm{FD}}$, corresponding to a~reduction of the effective dopant concentration ($N_{\mathrm{eff}}$), can be attributed to the acceptor compensation effect in $p$-type silicon under high $\gamma$~TID irradiation. This effect is associated with the formation of Silicon vacancy-Oxygen (VO (A-center)) \cite{hirata1966} as well as Boron-interstitial-Oxygen-interstitial  $\mathrm{B}_i\mathrm{O}_i$ \cite{Liao} deep-level trapping defects. When electrons are captured, negative space charge is introduced, thereby compensating 
and eventually over-compensating the initial acceptor concentration.

\begin{figure}
                \centering
                \includegraphics[width=0.7\linewidth]{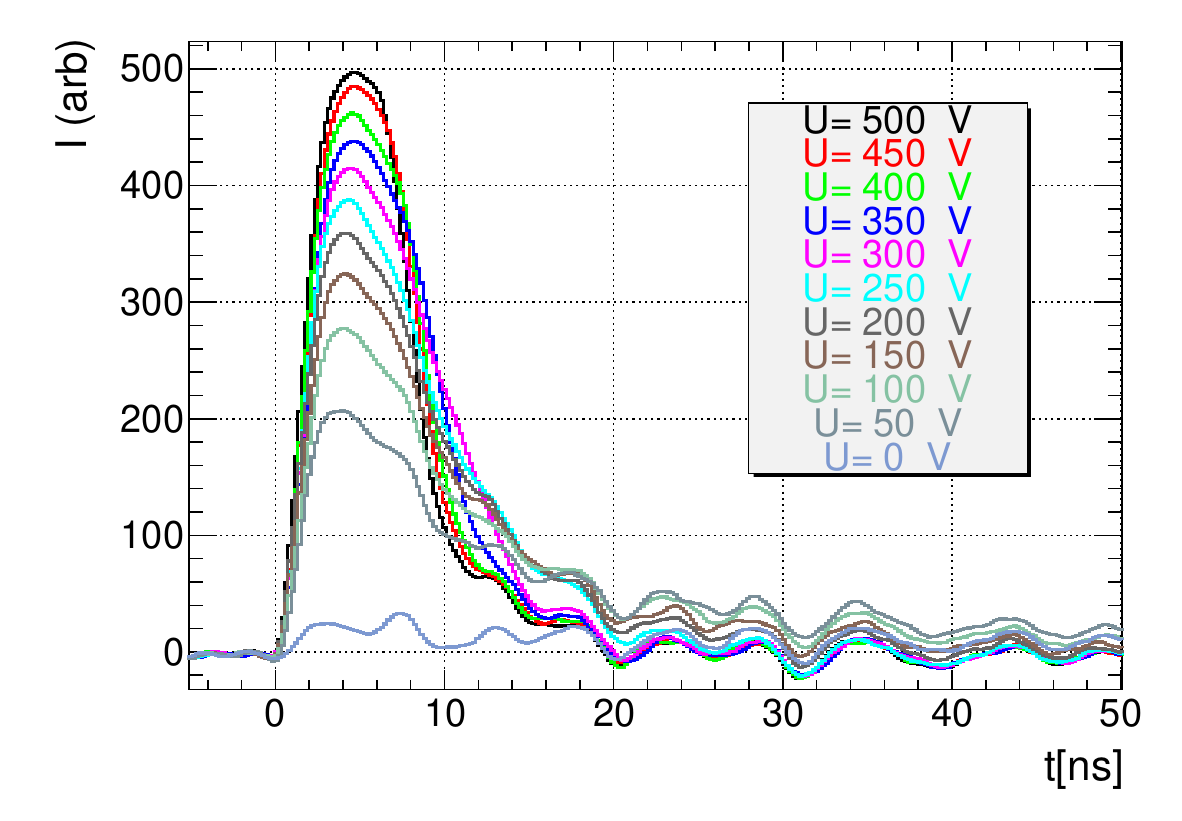}
                \caption{Top-TCT results for MD8 diode irradiated to 630~Mrad and subsequently annealed at 300\,$^\circ\mathrm{C}$.}
                   \label{fig:Fig8}
\end{figure}

Fig. \ref{fig:Fig8} shows the induced current pulses from a~TCT voltage scan between 0 and 500~V for an MD8 diode irradiated to 630~Mrad and subsequently annealed at 300\,$^\circ\mathrm{C}$. The strong electric field near the $n^{+}$ side results in a~high induced current immediately after the laser pulse at $t~=~0$, followed by a~decrease as carriers drift toward the $p^{+}$ side. This behavior confirms that after 300\,$^\circ\mathrm{C}$ annealing the bulk returns to $p$-type, indicating that space charge sign inversion is no longer present and the diode again exhibits an $n^{+}\text{-in-}p$ structure \cite{2024}.

\begin{figure}
                \centering
                \includegraphics[width=0.7\linewidth]{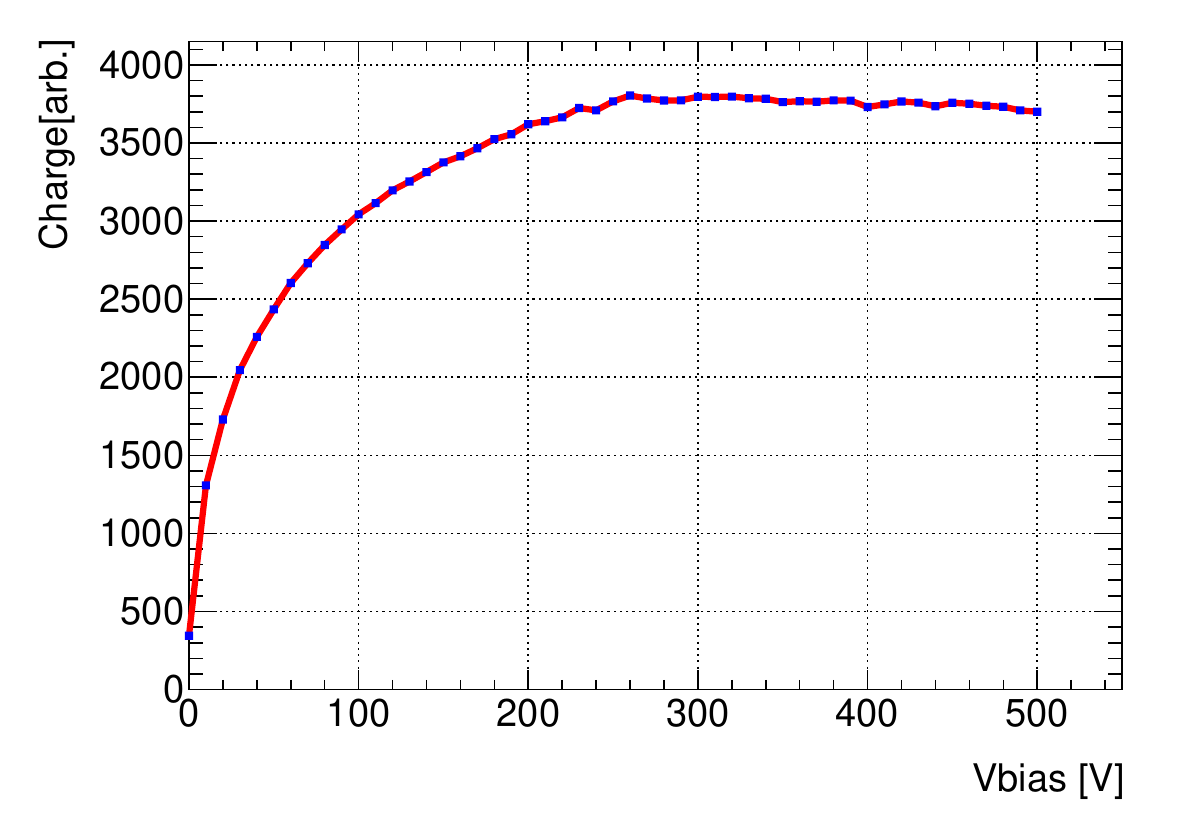}
                \caption{CCE obtained by integrating the transient current over the first 25~ns. $V_{\mathrm{FD}}$ = 260~V.}
                   \label{fig:Fig9}
\end{figure}

The $V_{\mathrm{FD}}$ can be estimated from the pulse shapes in Fig. \ref{fig:Fig8}; for fully depleted conditions, the current pulse shows a~sharp trailing edge, whereas below $V_{\mathrm{FD}}$ the signal approaches the baseline exponentially due to drift in the low electric field near the non-depleted bulk. The extracted $V_{\mathrm{FD}}$ lies between -250 and -300~V, in agreement with the CV measurements. The charge collection efficiency (CCE), obtained by integrating the transient current over the first 25~ns (Fig. \ref{fig:Fig9}), saturates at bias voltages around -260~V, consistent with the $V_{\mathrm{FD}}$ determined from both TCT and CV measurements.

The TCT results demonstrate that high temperature annealing effectively restores the electric field configuration of the diode after high TID irradiation. The recovery of $V_{\mathrm{FD}}$ and the re-establishment of $p$-type conductivity throughout the bulk indicate substantial annealing of radiation induced deep-level traps.

\subsection{Temperature dependence of leakage currents} \label{sec:Temperature dependence}

\begin{figure}
                                \centering
                \includegraphics[width=0.7\linewidth]{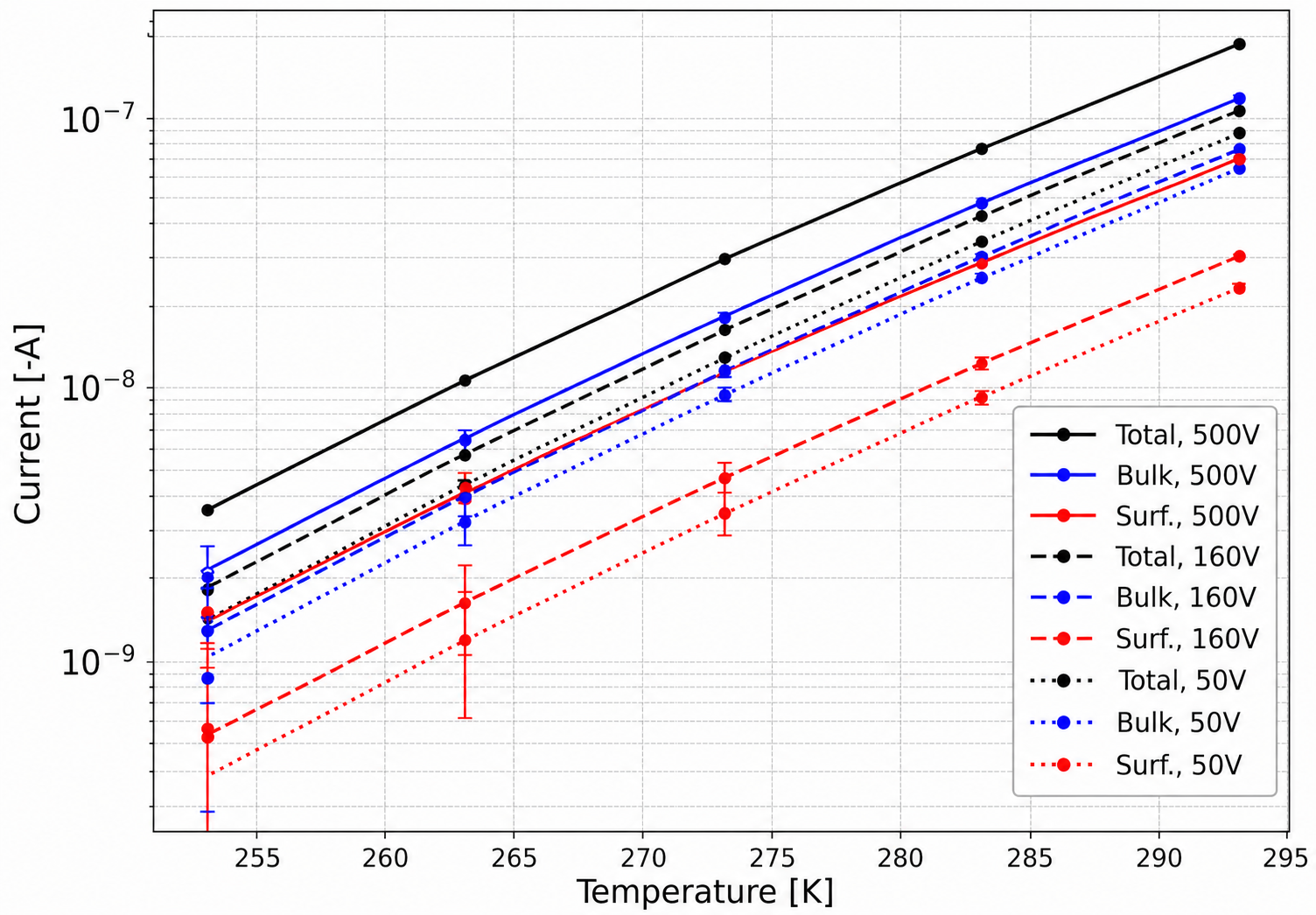}
                                \caption{Temperature dependence of total, bulk, and surface currents of mini irradiated to 100~krad and biased to -50~V, -160~V, and -500~V. Measured data are fitted with formula in Eq.~\eqref{eq:current_temperature}.}
                                      \label{fig:Fig12_new}
\end{figure}

The temperature dependence of bulk leakage currents in proton- and neutron-irradiated silicon sensors has been extensively studied and is summarized in Ref. \cite{Chilingarov}. In the present work, the temperature dependence of the total, surface, and bulk leakage currents was investigated for $\gamma$~irradiated minis exposed to a~dose of 100~krad. The leakage current was measured at a~fixed reverse bias while varying the temperature between \mbox{-20}\,$^\circ\mathrm{C}$ and \mbox{+20}\,$^\circ\mathrm{C}$. The measured currents were fitted using 
\begin{equation}
I(T) = A\,T^{2}\,\exp\!\left(-\frac{E_A}{2k_BT}\right),
\label{eq:current_temperature}
\end{equation}
where $k_B$ is the Boltzmann constant and $A$ and the activation energy $E_A$ were treated as free parameters for the total, surface, and bulk current components.
In depleted detectors, the bulk current originates from carrier generation in the depleted volume, while the surface current flows in the depleted inversion layer at the Si-SiO$_2$ interface. In both cases, electrons are the dominant carriers, leading to a~similar expected temperature dependence of the leakage current. Differences in the extracted activation energies for bulk and surface currents may arise from interface traps affecting the surface component. However, at the relatively low $\gamma$~doses considered here, no significant modification of radiation induced bulk defect levels is expected, and the bulk leakage current behavior should remain consistent with the standard generation model.
Fig. \ref{fig:Fig12_new} shows the bulk, surface, and total leakage currents plotted in Arrhenius representation, measured at bias voltages of -50~V, -160~V, and -~500~V. These three bias regions were selected to investigate the temperature dependence of the currents under different operational conditions: a low-bias region (-50~V), where the surface current contribution is expected to dominate; a region close to the full depletion voltage (-160~V); and an over-depleted region (-500~V). Table~\ref{tab:EA_vs_Vbias} summarizes the extracted activation energies for all three current components: total, bulk, and surface. No significant differences are observed between the activation energies derived from the temperature dependence of the bulk, surface, and total currents. The obtained values at bias voltage of -160~V are consistent with the result reported by Chilingarov, $E_A$ = (1.209 ± 0.007)~eV, for neutron- and proton-irradiated silicon detectors \cite{Chilingarov}. 
In addition, the activation energy of the bulk current measured in the mini sensor at bias voltage -160~V agree well with the results of the previous study on the MD8 diode irradiated to same TID of 100~krad, where a value of 1.208~eV was obtained at -150~V \cite{2025}.

\begin{table}[h]
\caption{Results of the fit of activation energies for total, bulk, and surface currents from temperature dependence of mini sensor irradiated to 100~krad and biased to -50~V, -160~V, and -~500~V.}
\centering
\begin{tabular}{c|c|c|c}
\hline
$V_{\text{bias}}$ [-V] & $E_A$ (tot) [eV] & $E_A$ (bulk) [eV] & $E_A$ (surf) [eV] \\
\hline
50  & 1.23 & 1.23 & 1.23 \\
160  & 1.20 & 1.21 & 1.20 \\
500 & 1.18 & 1.19 & 1.16 \\
\hline
\end{tabular}
\label{tab:EA_vs_Vbias}
\end{table}
\section{Conclusion}

Miniature strip sensors and MD8 diodes fabricated on ATLAS18 production wafers were irradiated with a~$^{60}$Co $\gamma$~source to TIDs ranging from 0.5 to 100~krad, relevant for the initial operating phase of the ATLAS ITk tracker, and additionally up to a few~Mrad to study the saturation of surface-related effects. The study combines low and high dose $\gamma$~irradiation effects, including results from our previous high dose $\gamma$~irradiation campaigns, to investigate surface- and bulk-related damage mechanisms and their annealing behaior.

Following irradiation, the total leakage current increases with increasing TID, while indications of surface current saturation are observed at $\approx2\,\mathrm{Mrad}$. For low TID irradiation levels relevant to ITk operation, the full depletion voltage $V_\mathrm{FD}$ remains unchanged after irradiation and annealing, indicating that the observed effects are predominantly surface-related and do not produce measurable bulk damage. High-temperature annealing above $\approx250\,^\circ\mathrm{C}$ restores the leakage currents close to their pre-irradiation values. Temperature-dependent measurements yield identical activation energies for bulk and surface currents, consistent with Chilingarov’s model, with $E_\mathrm{a} = (1.209 \pm 0.007)$ eV.

In contrast, at high $\gamma$~doses of several hundred Mrad, a~qualitatively different behavior is observed. The bulk leakage current increases linearly with TID, accompanied by a~strong reduction of $V_\mathrm{FD}$ from approximately $-275$ V to approximately $-20$ V, indicating significant space-charge redistribution caused by carrier trapping at radiation-induced deep levels. 
Annealing at 300\,$^\circ\mathrm{C}$ liberated trapped charges and returned $V_\mathrm{FD}$ close to its pre-irradiation value. However, residual deep-level defects may still have contributed to the high leakage current.

Overall, the results demonstrate that $\gamma$-induced effects in ATLAS18 strip sensors at TID levels relevant for the early phase of ITk operation remain limited and largely reversible, supporting stable strip detector performance during the initial operating period of the ATLAS ITk tracker. At the same time, the high-dose studies reveal previously unobserved effects induced by pure gamma irradiation, including trap-induced space-charge redistribution which could be removed by high temperature annealing.

\section*{Acknowledgments} \label{sec:aknowledgements}
This work was supported by the European Structural and Investment Funds and the Ministry of Education, Youth and Sports of the Czech Republic via projects LM2023040 CERN-CZ, and FORTE-CZ.02.01.01/00/22008/0004632, by the US Department of Energy, grant DE-SC0010107, of the Spanish R\&D grant PID2021-126327OB-C22, funded by MICIU/ AEI/10.13039/501100011033 and by ERDF/EU, by the Slovenian Research and Innovation Agency (research core funding No. P1-0135).

Copyright 2026 CERN for the benefit of the ATLAS Collaboration. Reproduction of this article or parts is allowed as specified in the CC-BY-4.0 license.


\end{document}